\documentclass[11pt,aps,pra,superscriptaddress,nofootinbib]{revtex4-2}
\usepackage{amsmath,amssymb,graphicx,bm}
\usepackage{orcidlink}
\usepackage{braket}
\usepackage{physics}
\usepackage{mathrsfs}
\usepackage{hyperref}

\begin{document}

\title{Quantum Optical Simulator for Unruh–DeWitt Detector Dynamics}

\author{Tai Hyun Yoon\,\orcidlink{0000-0002-2408-9295}}
\email{thyoon@korea.ac.kr}
\affiliation{Department of Physics, Korea University, Anam-ro 145, Seongbuk-gu, Seoul 02841, Republic of Korea}
\affiliation{Center for Molecular Spectroscopy and Dynamics, IBS, Anam-ro 145, Seongbuk-gu, Seoul 02841, Republic of Korea}

\begin{abstract}
We present a quantum-optical platform based on coherently seeded entangled nonlinear biphoton sources (ENBSs) that realizes an operational analog of detector--environment interactions inspired by the Unruh--DeWitt (UDW) model. 
The system is implemented using phase-controlled single-photon frequency-comb (SPFC) sources, enabling precise control of amplitude, phase, and temporal structure of correlated photon pairs. Within this framework, the signal mode acts as an effective probe, while the idler mode and associated photonic modes form a controllable environment.
We derive the effective interaction Hamiltonian and corresponding Lindblad master equation for two coherently seeded ENBS units, and obtain analytical expressions for the signal photon number \(N_{\mathrm{sig}}(t)\) and second-order correlation function \(g^{(2)}(0;t)\). 
We show that phase-dependent detector response arises from coherent superposition of multiple excitation pathways. 
In particular, single-source observables depend only on local pump--seed phase parameters, while phase dependence associated with the global interference phase emerges exclusively when multiple sources are coherently combined. 
The resulting dynamics enable phase-resolved control of intensity correlations and reduced-state properties, including fidelity and coherence visibility. 
Our results demonstrate that coherently seeded ENBS systems provide a versatile and experimentally accessible platform for investigating phase-controlled quantum correlations and detector--environment interactions in a photonic setting. 
While the present system does not reproduce continuum quantum field dynamics, causal structure, or thermal effects associated with accelerated detectors, it captures the operational role of environment-induced correlations in determining observable detector response.
\end{abstract}

\maketitle

\section{Introduction}

Understanding the interaction between localized quantum systems and quantum fields lies at the foundation of relativistic quantum information and quantum field theory in curved spacetime. 
The Unruh--DeWitt (UDW) detector model provides a minimal and powerful framework in which a two-level system probes quantum field fluctuations along a prescribed trajectory, with its excitation probability determined by the field correlation function evaluated along the detector worldline \cite{Unruh1976,DeWitt1979,Birrell1982}. 
This framework has enabled profound insights into phenomena such as the Unruh effect, Hawking radiation, and the structure of quantum correlations in relativistic settings, including entanglement generation and correlation harvesting in quantum fields \cite{Reznik2003,Pozas2015,Salton2015,Simidzija2018}. Recent theoretical developments have further extended the UDW framework to include localized wavepacket detection, detectors in superpositions of trajectories, and refined descriptions of detector--field interactions in quantum field theory \cite{Martinez2013,Foo2020,Gale2023,Perche2024}.

Despite its conceptual importance, direct experimental access to UDW-type detector--field dynamics remains challenging. 
In particular, hallmark features of relativistic quantum field theory---including continuum field modes, causal structure, and thermal (Kubo--Martin--Schwinger, KMS) behavior associated with acceleration---are difficult to reproduce in controlled laboratory systems. 
As a result, considerable effort has been devoted to developing analog quantum simulators that capture selected aspects of detector--field interactions in experimentally accessible platforms, including nonlinear optical systems, superconducting circuits, and ultracold atomic gases \cite{Adjei2020,Hu2019,Viermann2022,Wilson2011,Laguna2017,Sheng2021}.

In this work, we introduce a quantum-optical platform based on coherently seeded entangled nonlinear biphoton sources (ENBSs) that realizes an \emph{operational analog of detector--environment interactions inspired by the UDW model}. 
Our system employs phase-coherent parametric processes driven by optical frequency combs, building on established techniques in nonlinear optics and multi-photon interference \cite{Boyd2008,Ou1990,Ling2008}, enabling precise control of amplitude, phase, and temporal structure of correlated photon pairs. 
Within this framework, the signal mode plays the role of an effective detector degree of freedom, while the idler mode and vacuum fluctuations constitute a controllable photonic environment.

The key feature of this platform is the ability to engineer and probe phase-resolved correlations between the effective detector and its environment. 
By coherently seeding the idler modes and controlling relative phases across multiple sources, we realize a tunable interference landscape in which detector response, quantified through photon number and second-order correlation functions, is governed by the coherence properties of the underlying quantum state. 
This enables a direct investigation of how distinguishability, coherence, and environment-induced correlations shape observable detector signals.

We emphasize that the present platform does not constitute a full simulation of relativistic quantum field dynamics. 
In particular, it does not reproduce continuum field correlators, causal horizon structure, or KMS thermal behavior associated with uniformly accelerated observers. 
Rather, it provides a controllable and experimentally accessible realization of the \emph{operational structure} of detector--environment coupling central to the UDW framework, including the role of correlation functions and phase-dependent response. This approach is complementary to relativistic quantum information protocols explored in quantum optics and superconducting platforms \cite{Friis2013,Ahmadi2014,Bruschi2016}.

Within this operational perspective, our approach connects naturally to developments in quantum optics and quantum information, where interference, complementarity, and decoherence are understood in terms of system--environment correlations \cite{Maccone2015,Herzog1995,Heuer2015,Qian2020}. In particular, the ENBS platform enables quantitative access to reduced-state properties such as coherence visibility and fidelity, which encode the degree of distinguishability between environmental states and directly influence observable interference.

Using this framework, we derive analytic expressions for the signal photon number $N_{\mathrm{sig}}(t)$ and the second-order correlation function $g^{(2)}(0;t)$, explicitly identifying the role of phase-controlled interference and environment-induced correlations. 
We show that phase-dependent detector response emerges from coherent superposition of multiple interaction pathways, while single-source observables remain phase-independent. 
Furthermore, we analyze the reduced signal-mode density matrix and quantify its coherence and correlation properties through fidelity and related measures.

Our results establish coherently seeded ENBS systems as a versatile platform for exploring detector--environment physics in a fully quantum-optical setting. 
Beyond their connection to UDW-inspired models and related analog simulation approaches \cite{Adjei2020}, these systems provide a powerful testbed for investigating the interplay between coherence, distinguishability, and measurement in photonic quantum systems, with potential applications in quantum simulation, quantum metrology, and fundamental studies of quantum correlations.

\begin{figure*}[t]
    \centering
    \includegraphics[width=0.8\linewidth]{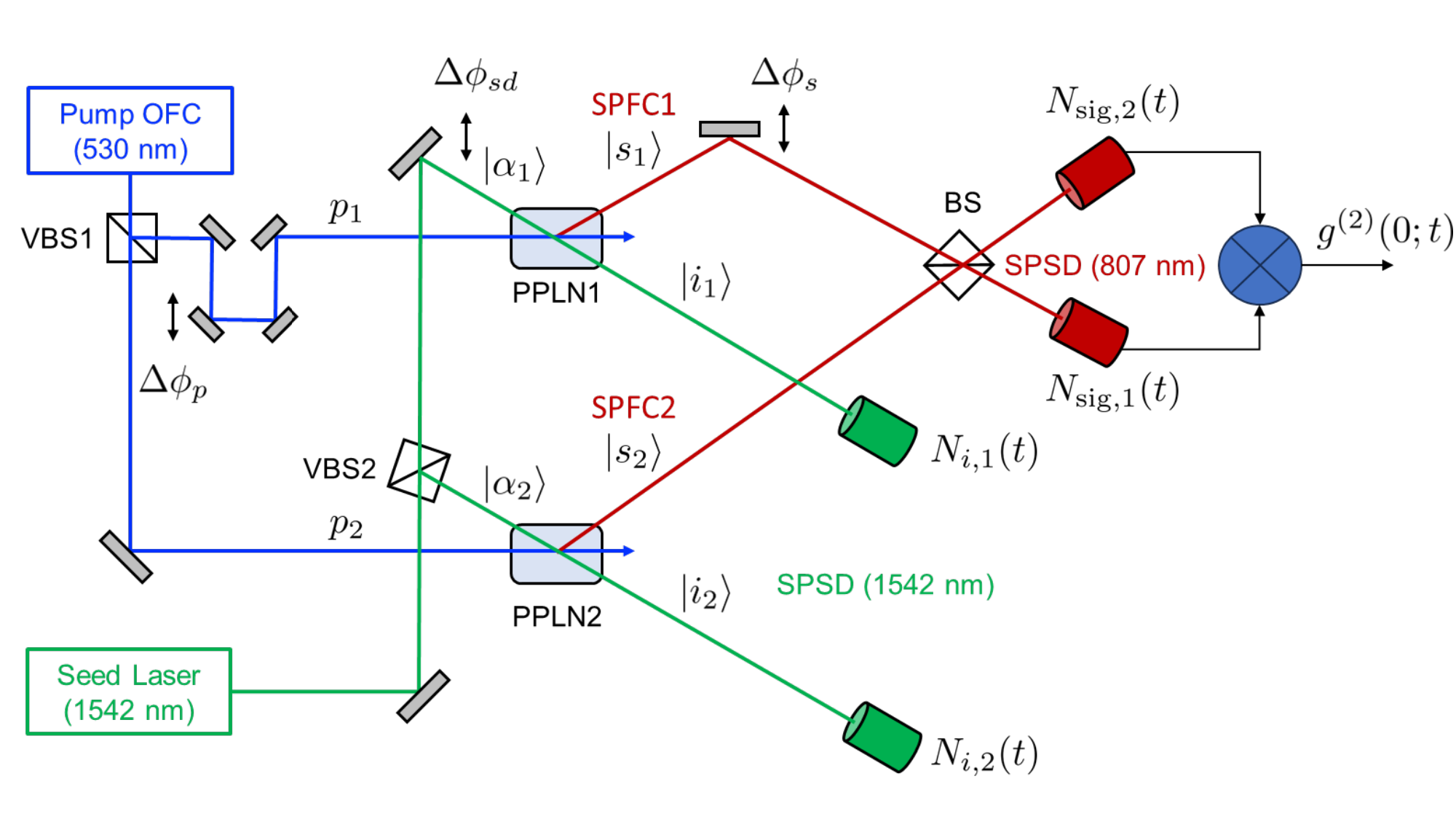}
   \caption{\textbf{Quantum-optical platform for phase-resolved detector--environment analog dynamics}. 
Two entangled nonlinear biphoton sources (ENBSs), each realized using a seeded single-photon frequency-comb (SPFC) generator, are implemented with periodically poled lithium niobate (PPLN) crystals pumped by optical frequency combs. 
Coherent seed fields at 1542~nm with a controllable relative phase \( \Delta\phi_{\text{sd}} \) are injected into the idler paths. 
The resulting signal outputs at 807~nm exhibit phase-dependent quantum correlations arising from coherent parametric interactions. 
The mean photon number \( N_{\rm sig}(t) \) is measured by removing the beam splitter (BS), while the second-order correlation function \( g^{(2)}(0;t) \) is obtained via coincidence detection at the two outputs. 
Alternatively, a single detector placed after the combining BS measures first-order coherence via the phase-dependent detection rate \( R(\Delta\phi_{\rm sd}) \). 
BS: beam splitter; VBS: variable beam splitter; SPSD: single-photon-sensitive detector.}
    \label{Fig1}
\end{figure*}

Figure~\ref{Fig1} illustrates the architecture of the quantum-optical platform~\cite{Lee2018,Yoon2021}. 
Two SPFC sources are pumped by optical frequency combs at 530~nm, and their idler channels are seeded with coherent fields \( \alpha_1 \) and \( \alpha_2 \) at 1542~nm, with tunable relative phase \( \Delta\phi_{\text{sd}} \). 
In addition to the seeding phase, the system involves independent phase degrees of freedom associated with the pump and signal paths, denoted by \( \Delta\phi_p \) and \( \Delta\phi_s \), respectively. 
Here, \( \Delta\phi_p \) represents the relative phase between the pump fields driving the two nonlinear interactions, while \( \Delta\phi_s \) accounts for the phase accumulated along the signal paths, including contributions from propagation and the combining beam splitter. These phase contributions combine to define an effective interferometric phase 
\(
\Phi = \Delta\phi_p - (\Delta\phi_{\text{sd}} + \Delta\phi_s),
\)
which governs the interference between distinct excitation pathways.

The resulting signal photons at 807~nm exhibit controllable quantum correlations governed by these phase differences. 
In particular, observable interference effects arise from coherent superposition of excitation pathways and depend on an effective phase combination determined by the interplay of pump, seeding, and signal phases. 
This enables phase-resolved probing of detector--environment interactions in an engineered photonic setting \cite{Breuer2002,Rivas2014}.

\textit{Different detection configurations access complementary observables.}
Removing the combining beam splitter (BS) yields the mean signal photon number \( N_{\rm sig}(t) \); coincidence detection at the BS outputs provides the second-order correlation function \( g^{(2)}(0;t) \); and monitoring a single output after the BS measures the phase-dependent single-photon detection rate \( R(\Delta\phi_{\rm sd}) \), which reflects first-order coherence of the signal field~\cite{Lee2018,Yoon2021}. 
This coherence arises from interference between multiple excitation pathways and encodes the degree of distinguishability of the underlying quantum states.

Compared to analog gravity platforms based on Bose--Einstein condensates (BECs)~\cite{Hu2019,Viermann2022,Laguna2017,Sheng2021}, which probe continuum quantum fields and can exhibit thermal signatures associated with effective horizons, the present photonic platform offers complementary capabilities. 
Specifically, it provides precise phase control, direct optical access to correlation functions, and scalability within well-established nonlinear optical architectures.

In contrast to continuum-field simulators, our approach focuses on the operational structure of detector response: namely, how measurable observables depend on correlations between a localized probe (signal mode) and its environment (idler mode and vacuum fluctuations). 
This enables controlled exploration of phase-dependent detector response and environment-induced coherence in a fully quantum-optical setting.

We derive the Hamiltonian dynamics and corresponding master equation governing the ENBS system, leading to analytical expressions for the signal photon number \( N_{\text{sig}}(t) \) and the second-order correlation function \( g^{(2)}(0;t) \). 
We show that phase dependence arises from coherent superposition of multiple interaction pathways, while single-source contributions remain phase-independent. 
The resulting observables provide direct access to correlation properties of the system and their dependence on controllable experimental parameters.

Beyond intensity-based observables, the platform enables quantitative characterization of reduced-state properties such as coherence visibility and fidelity, which encode the distinguishability of environmental states and directly influence interference behavior. 
These quantities provide a natural bridge between quantum optical observables and information-theoretic measures of system--environment correlations.

Although the present system employs discrete optical modes rather than a field-theoretic continuum, it captures the essential operational feature that detector response is governed by correlations with an environment. 
This establishes the ENBS platform as a versatile and experimentally accessible testbed for investigating detector--environment physics and phase-resolved quantum correlations in photonic systems.

\begin{table}[t]
\centering
\caption{\textbf{Operational correspondence between the ENBS platform and elements of the Unruh--DeWitt (UDW) detector framework.}
The mapping highlights analogies at the level of detector response and correlation structure, rather than a full equivalence to relativistic quantum field dynamics.}
\small
\setlength{\tabcolsep}{4pt}
\renewcommand{\arraystretch}{1.15}
\begin{tabular}{p{0.32\textwidth}|p{0.32\textwidth}|p{0.32\textwidth}}
\hline
\textbf{UDW Detector Concept} & \textbf{ENBS Optical Platform} & \textbf{Interpretation} \\
\hline

Localized detector (two-level system) 
& Signal mode 
& Effective probe of the system response \\

Quantum field (continuum modes)
& Idler mode + photonic environment 
& Tunable environment controlling correlations \\

Detector--field interaction 
& Nonlinear parametric coupling 
& Source of correlated excitation processes \\

Detector response function 
& Signal photon number \(N_{\mathrm{sig}}(t)\) 
& Observable excitation dynamics \\

Field correlation function 
& Phase-dependent coherence and intensity correlations 
& Determines interference and correlation structure \\

Environmental decoherence 
& Tracing over idler modes 
& Reduced-state loss of coherence \\

Interference between pathways 
& Phase-controlled excitation pathways 
& Origin of phase-dependent detector response \\

\hline
\end{tabular}
\label{tab:UDW_mapping}
\end{table}

\section{Quantum Optical Simulator Framework: Modeling Detector--Environment Dynamics}

We consider a quantum-optical platform shown in Fig.~\ref{Fig1} based on coherently seeded entangled nonlinear biphoton sources (ENBSs) \cite{Lee2018,Yoon2021}, in which signal and idler modes are generated via parametric interactions driven by phase-coherent pump fields. 
The system provides a controllable setting to investigate detector--environment interactions in an operational framework inspired by the Unruh--DeWitt (UDW) model. Such detector-based approaches provide an operational framework for probing quantum fields through localized interactions \cite{Martinez2013,Perche2024}.

\subsection{Model and assumptions}

Each ENBS is described by an effective interaction Hamiltonian of the form
\begin{equation}
\hat{H}_{\mathrm{int}} = i\hbar \left( g \, e^{i\phi_p} \hat{a}_s^\dagger \hat{a}_i^\dagger - g^* \, e^{-i\phi_p} \hat{a}_s \hat{a}_i \right),
\end{equation}
where \( \hat{a}_s \) and \( \hat{a}_i \) denote signal and idler annihilation operators, respectively, and \( g \) is the effective nonlinear coupling strength. 
The phase \( \phi_p \) represents the pump phase.

The idler modes are coherently seeded with amplitudes \( \alpha_1 \) and \( \alpha_2 \), introducing a controllable relative phase \( \Delta\phi_{\text{sd}} \). 
The signal modes propagate through interferometric paths that introduce an additional phase difference \( \Delta\phi_s \), including contributions from propagation and beam splitter mixing.

We focus on the weak-gain regime, where the average photon number remains much smaller than unity. 
In this regime, higher-order multiphoton processes and time-ordering corrections are negligible, and the dynamics are well described within a truncated Hilbert space.

\subsection{Phase structure and signal photon number}

The ENBS system contains three independent phase degrees of freedom,
\begin{equation}
\Delta\phi_p, \quad \Delta\phi_{\mathrm{sd}}, \quad \Delta\phi_s,
\end{equation}
corresponding to the pump phase, idler seeding phase, and signal-path phase, respectively. 
These phases combine into two physically distinct phase variables that govern different aspects of the dynamics.

\textit{(i) Local phase (single-source dynamics).}  
For each ENBS, the signal photon number is governed by a local pump--seed phase
\begin{equation}
\Phi_{N,j} = \phi_{p,j} - \phi_{\alpha,j},
\end{equation}
which determines the phase of the seeded parametric amplification process. 
In the presence of coherent idler seeding, the single-source signal photon number takes the form
\begin{equation}
N_{\mathrm{sig},j}(t)
= \left[ 1 + |\alpha_j|^2 + 2|\alpha_j|\cos(\Phi_{N,j}) \right]\sinh^2(gt),
\end{equation}
showing explicit dependence on the local phase \( \Phi_{N,j} \).
This phase dependence arises from interference between the vacuum and seeded excitation pathways within a single nonlinear interaction.
The resulting phase-resolved amplification dynamics are illustrated in Supplementary Fig.~S1.

\textit{(ii) Global phase (two-source interference).}  
When two ENBS units are coherently combined, interference between distinct excitation pathways introduces an additional phase degree of freedom
\begin{equation}
\Phi = \Delta\phi_p - (\Delta\phi_{\mathrm{sd}} + \Delta\phi_s),
\end{equation}
which governs interference between the two sources. 
In this case, the total signal photon number detected at a selected output port of the combining beam splitter (BS) in Fig.~1 takes the form
\begin{equation}
N_{\mathrm{sig}}(t)
= N_1(t) + N_2(t) + 2\sqrt{N_1(t)N_2(t)} \cos \Phi,
\end{equation}
where \( N_1(t) \) and \( N_2(t) \) denote the individual source contributions.

Here, \( N_{\mathrm{sig}}(t) \) corresponds to the photon number measured at one output port of the BS, and the signal-path phase shift \( \Delta\phi_s \), introduced by translating the BS, serves as an experimental control parameter that directly tunes the interference phase \( \Phi \).
The distinct roles of the local phase \( \Phi_{N,j} \) and the global phase \( \Phi \) are further illustrated by the two-dimensional correlation structure shown in Supplementary Fig.~S2.

Thus, the ENBS platform exhibits a clear separation of phase roles:  
the local phase \( \Phi_{N,j} \) governs single-source amplification dynamics, while the global phase \( \Phi \) governs interference between distinct excitation pathways.

\subsection{Second-order correlation function}

The normalized second-order correlation function is defined as
\begin{equation}
g^{(2)}(0;t) = 
\frac{\left\langle 
\hat{a}_s^\dagger(t)\hat{a}_s^\dagger(t)\hat{a}_s(t)\hat{a}_s(t) 
\right\rangle}
{\left\langle 
\hat{a}_s^\dagger(t)\hat{a}_s(t) 
\right\rangle^2}.
\end{equation}

This quantity characterizes intensity correlations in the signal mode and is sensitive to both spontaneous and stimulated processes. 
We emphasize that \( g^{(2)} \) does not constitute a direct entanglement witness; in particular, large values may arise in the low-count regime due to normalization by \( \langle n \rangle^2 \). 
Accordingly, it is interpreted here as a probe of correlation structure rather than a measure of entanglement.

Building on the phase structure introduced in Sec.~II.B, the second-order correlation function provides a direct probe of interference between distinct excitation pathways. 
For a single ENBS, no global interference occurs, and the correlation function is determined solely by local amplification dynamics. 
In contrast, when two ENBS units are coherently combined, \( g^{(2)}(0;t) \) acquires a pronounced dependence on the global interference phase \( \Phi \).

In particular, the phase-sensitive contribution to the correlation function scales as \( \sin^2\Phi \), reflecting interference between indistinguishable two-photon excitation pathways. 
This behavior contrasts with the signal photon number, which contains an interference term proportional to \( \cos\Phi \), and highlights the distinct roles of first- and second-order observables in probing phase-dependent correlations (see Supplementary Note II and Fig.~S2).

\subsection{Reduced-state description and fidelity}

To quantify coherence and distinguishability, we consider the reduced signal-mode density matrix obtained by tracing over the idler degrees of freedom. 
The overlap between conditional signal states is characterized by the fidelity
\begin{equation}
F = \left| \langle \alpha_1,1 \mid \alpha_2,1 \rangle \right|,
\end{equation}
where the normalized single-photon-added coherent states (SPACS) are defined as
\begin{equation}
|\alpha_j,1\rangle
=
\frac{\hat{a}^\dagger |\alpha_j\rangle}{\sqrt{1+|\alpha_j|^2}},
\end{equation}
following Refs.~\cite{Agarwal1991,Zavatta2004}.

Using this definition, the fidelity can be written as
\begin{equation}
F
=
\frac{\left|1+\alpha_1^*\alpha_2\right|}
{\sqrt{(1+|\alpha_1|^2)(1+|\alpha_2|^2)}}
\exp\!\left[
-\frac{|\alpha_1|^2}{2}
-\frac{|\alpha_2|^2}{2}
+\mathrm{Re}(\alpha_1^*\alpha_2)
\right].
\end{equation}

Writing
\(
\alpha_j = |\alpha_j| e^{i\phi_j}
\)
with \( \Delta\phi_{sd} = \phi_2-\phi_1 \), one obtains
\(
\alpha_1^*\alpha_2 = |\alpha_1||\alpha_2|e^{i\Delta\phi_{sd}}
\),
so that
\begin{equation}
F
=
\frac{\left|1+|\alpha_1||\alpha_2|e^{i\Delta\phi_{sd}}\right|}
{\sqrt{(1+|\alpha_1|^2)(1+|\alpha_2|^2)}}
\exp\!\left[
-\frac{|\alpha_1|^2}{2}
-\frac{|\alpha_2|^2}{2}
+|\alpha_1||\alpha_2|\cos\Delta\phi_{sd}
\right].
\end{equation}

For symmetric seeding, \( |\alpha_1|=|\alpha_2|=|\alpha| \), this reduces to
\begin{equation}
F
=
\frac{\left|1+|\alpha|^2 e^{i\Delta\phi_{sd}}\right|}{1+|\alpha|^2}
\exp\!\left[
-|\alpha|^2\bigl(1-\cos\Delta\phi_{sd}\bigr)
\right].
\end{equation}

The fidelity depends on both amplitude and phase differences between the coherent states and approaches unity only when the two states become indistinguishable in phase space, thereby directly determining the interference visibility and coherence of the signal mode.

\begin{figure}[tbh!]
    \centering
    \includegraphics[width=0.9\linewidth]{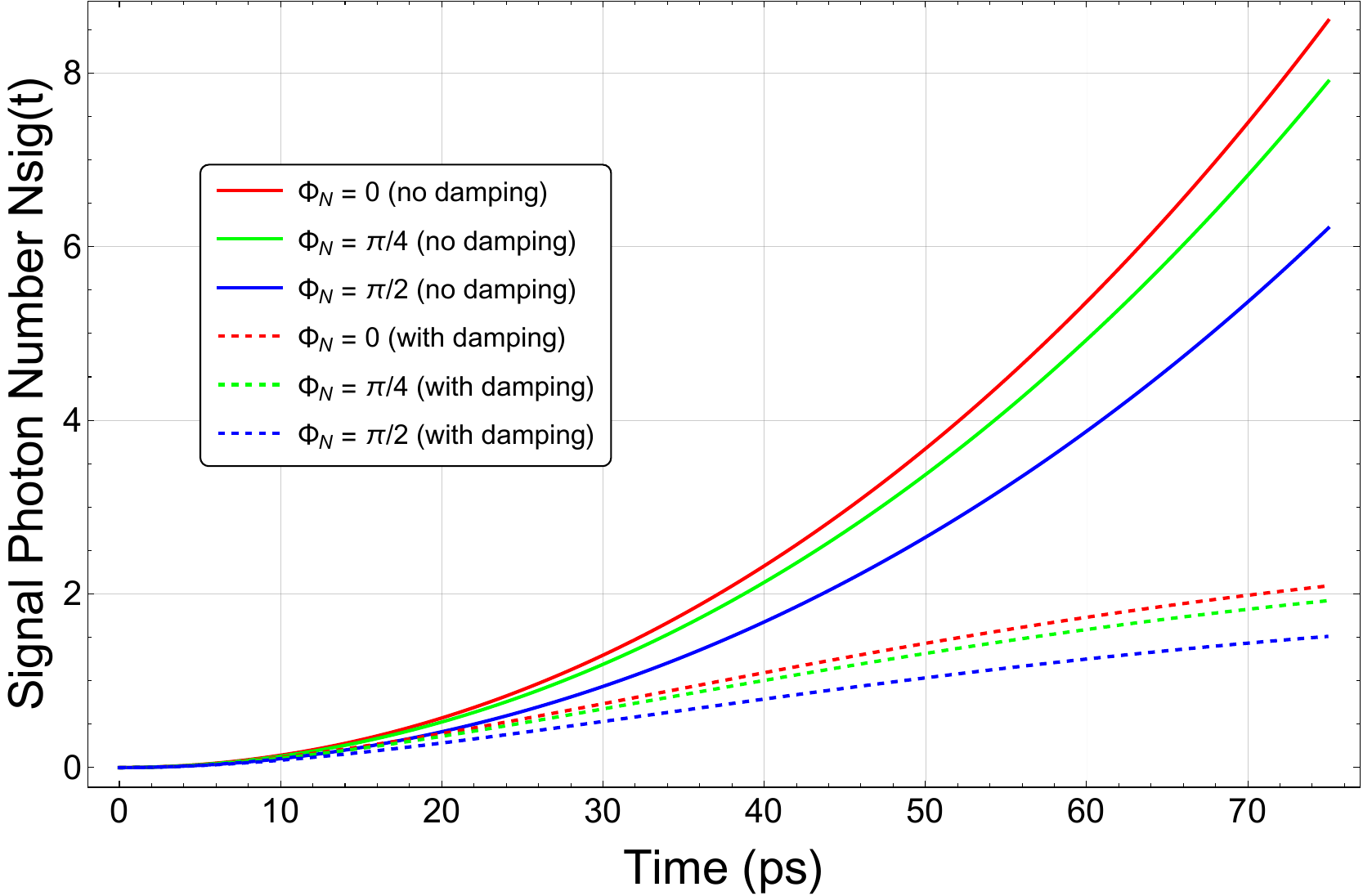}
\caption{\textbf{Time evolution of the signal photon number under coherent seeding and dissipation.}
(a) Growth of the signal photon number \( N_{\mathrm{sig}}(t) \) in the absence of damping, showing exponential amplification due to parametric interaction. 
(b) Saturation behavior in the presence of signal loss, illustrating the balance between gain and dissipation. 
The dynamics reflect the buildup of excitation in the signal mode under coherent pumping, with phase dependence emerging only when interference between multiple sources is present.}
    \label{Fig2}
\end{figure}

\subsection{Operational interpretation}

Within this framework, the signal mode acts as an effective probe whose response is governed by correlations with an environment formed by the idler mode and vacuum fluctuations. 
The phase-dependent observables derived above reflect the structure of these correlations.

This establishes an operational analogy to detector--environment interactions in the UDW model, in which detector response is determined by field correlation functions. 
However, the present system does not reproduce continuum field dynamics, causal structure, or thermal effects associated with accelerated detectors, and should therefore be interpreted as a controllable quantum-optical analog rather than a full relativistic field simulation.

\section{Phase-Resolved Correlations and Quantum Information Signatures}

We now analyze the behavior of experimentally accessible observables in the ENBS platform, focusing on phase-resolved signal dynamics, correlation functions, and quantum information measures. 
All quantities are evaluated using the framework introduced in the previous section, without rederiving expressions.

\subsection{Signal dynamics and detector response}

Figure~\ref{Fig2} shows the time evolution of the signal photon number \( N_{\mathrm{sig}}(t) \) under coherent seeding. 
In the absence of damping, the signal exhibits exponential growth due to parametric amplification, reflecting the coherent buildup of excitation in the signal mode driven by the nonlinear interaction. 
Including loss leads to saturation, indicating a steady-state regime set by the competition between amplification and dissipation.

This behavior defines a finite effective interaction window, during which the signal mode accumulates excitation before reaching a balance between gain and loss. 
Within the detector analogy, this corresponds to the temporal profile of the effective detector response governed by its coupling to the environment.

We emphasize that the phase dependence of \( N_{\mathrm{sig}}(t) \) arises only in the presence of interference between multiple sources. 
For a single source, the photon number depends on the local phase 
\( \Phi_{N,j} \), which governs the seeded amplification process. 
However, no dependence on the global interference phase \( \Phi \) arises 
in the absence of coherent combination between multiple sources.

\begin{figure}[tbh!]
    \centering
    \includegraphics[width=0.9\linewidth]{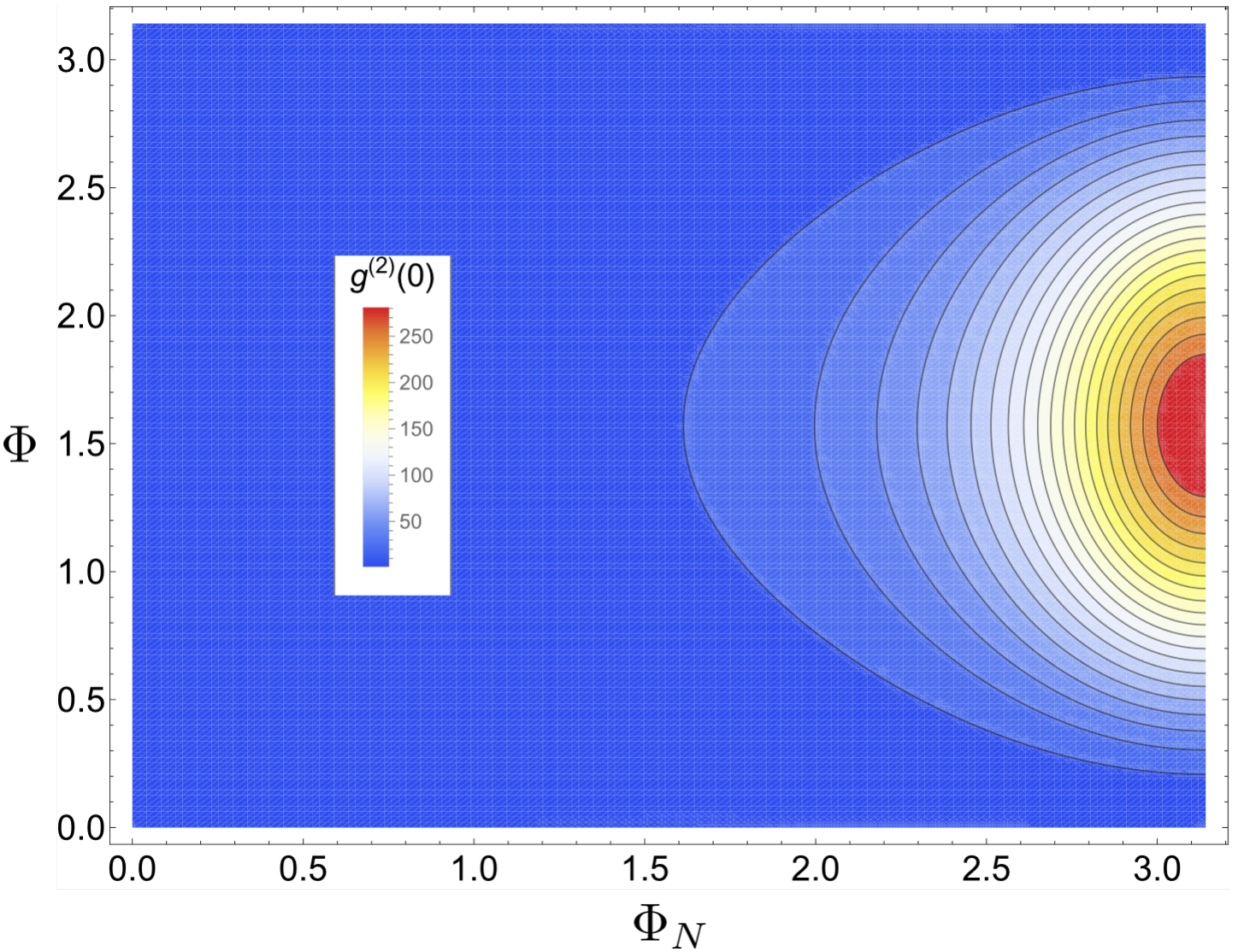}
\caption{\textbf{Phase dependence of the second-order correlation function \( g^{(2)}(0) \).}
Color map of \( g^{(2)}(0) \) as a function of the phase parameters \( \Phi \) and \( \Phi_N \), showing strong phase-dependent modulation of intensity correlations. 
A pronounced enhancement of \( g^{(2)}(0) \) is observed near specific phase combinations, corresponding to constructive interference between excitation pathways. 
In contrast, regions of low \( g^{(2)}(0) \) indicate reduced correlation strength. 
The second-order correlation function reflects normalized intensity correlations arising from interference effects and should not be interpreted as a direct measure of entanglement, as large values can occur in regimes of low mean photon number due to normalization.}
    \label{Fig3}
\end{figure}

\subsection{Phase-controlled correlations}

The second-order correlation function \( g^{(2)}(0;t) \), shown in Fig.~\ref{Fig3}, exhibits strong dependence on the global phase \( \Phi \). 
This phase dependence originates from interference between indistinguishable excitation pathways associated with the two coherently seeded sources, which modulates the joint detection probability of signal photons. 
Maximal bunching occurs near \( \Phi = \pi/2 \), where these pathways interfere constructively in the correlation signal, enhancing intensity fluctuations relative to the mean photon number.

Importantly, \( g^{(2)}(0;t) \) reflects normalized intensity correlations and does not constitute a direct entanglement witness. 
In particular, large values can arise in regimes of low mean photon number due to normalization by \( \langle n \rangle^2 \), even in the absence of strong quantum correlations. 
Thus, \( g^{(2)} \) should be interpreted as a probe of correlation structure arising from interference and statistical fluctuations, rather than a direct measure of entanglement.

The observed phase dependence demonstrates that correlation properties can be controlled independently of the average signal intensity through phase tuning. 
Within the detector analogy, this highlights how interference between excitation pathways in the environment can shape the measured detector response without necessarily altering the mean excitation probability.

\begin{figure}[t]
\centering
\includegraphics[width=0.9\linewidth]{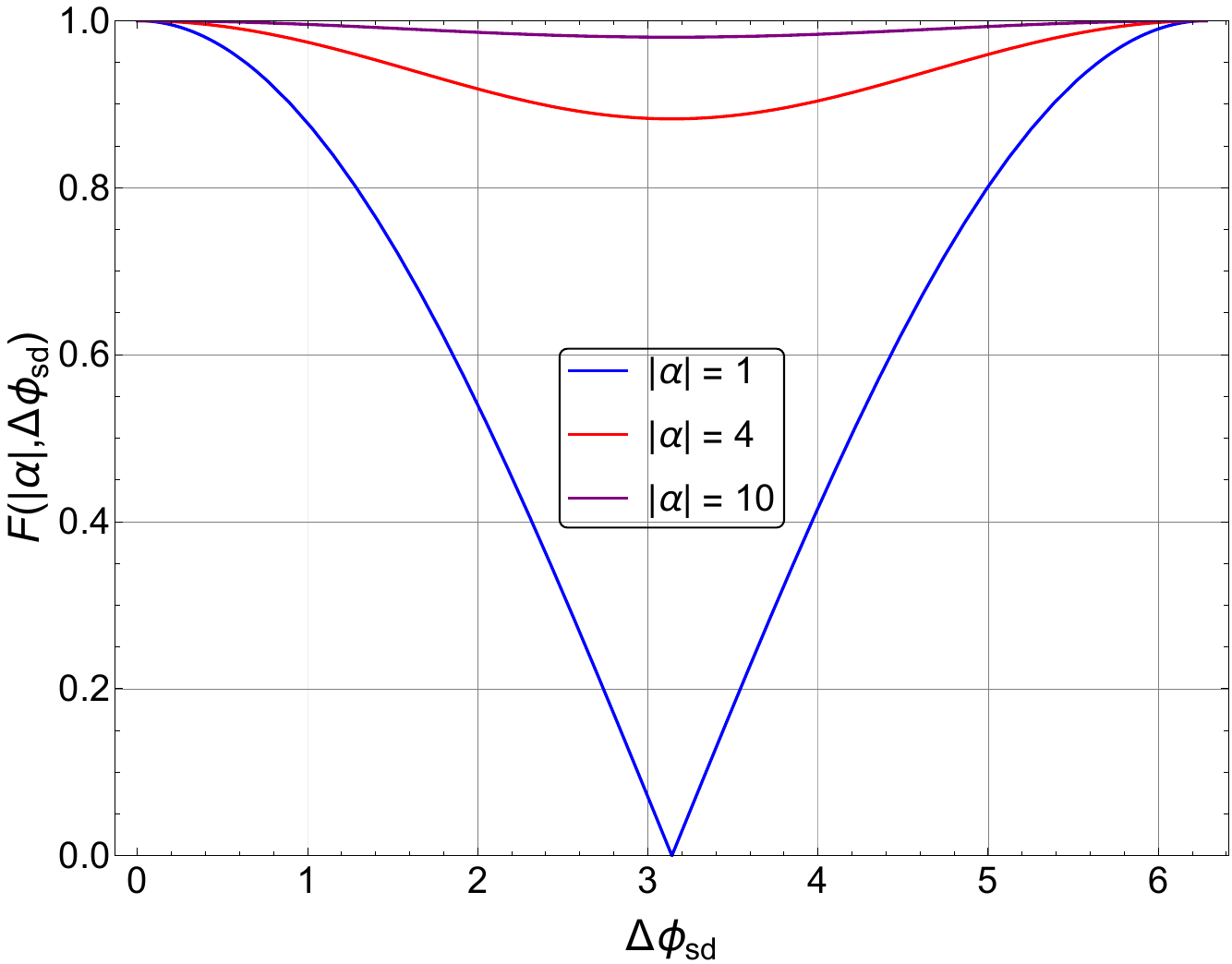}
\caption{\textbf{Fidelity of conditional signal states as a function of seeding phase difference and amplitude.}
The fidelity \(F = |\langle \alpha_1,1 | \alpha_2,1 \rangle|\) is shown as a function of the relative seeding phase \( \Delta\phi_{sd} \) for different coherent amplitudes \( |\alpha| \), assuming symmetric seeding \( |\alpha_1| = |\alpha_2| = |\alpha| \). 
For small amplitudes, the fidelity exhibits pronounced phase dependence, reflecting quantum interference between distinguishable conditional states. 
As the amplitude increases, the fidelity remains close to unity over a broad range of \( \Delta\phi_{sd} \), indicating reduced sensitivity to phase differences due to the increasingly classical character of the underlying coherent states.}
\label{Fig4}
\end{figure}

\subsection{Fidelity and coherence}

We now examine the phase dependence of the fidelity \(F\), defined in the previous section, which quantifies the distinguishability of conditional signal states associated with different idler configurations. 
Figure~\ref{Fig4} shows \(F(\Delta\phi_{\mathrm{sd}})\) for different seeding amplitudes. 
The fidelity reaches a maximum near \( \Delta\phi_{\mathrm{sd}} = 0 \), where the corresponding conditional states are most indistinguishable, and decreases as the phase difference increases, reflecting increasing distinguishability in the effective environment.

This behavior is governed by the phase-dependent overlap of the underlying single-photon-added coherent states, which determines the degree of coherence retained in the reduced signal mode.

For large amplitudes, the fidelity remains close to unity over a broad range of \( \Delta\phi_{\mathrm{sd}} \), reflecting reduced sensitivity to phase differences as the coherent states become increasingly classical. 
Conversely, for small amplitudes, the fidelity exhibits stronger phase dependence, indicating enhanced sensitivity to phase and more pronounced quantum interference effects.

These results show that phase control provides a direct means of tuning the distinguishability of conditional states and the coherence properties of the signal mode. 
Within the detector analogy, this corresponds to controlling how strongly the environment encodes which-path information, thereby governing the balance between coherence and distinguishability.

\begin{figure}[t]
\centering
\includegraphics[width=\linewidth]{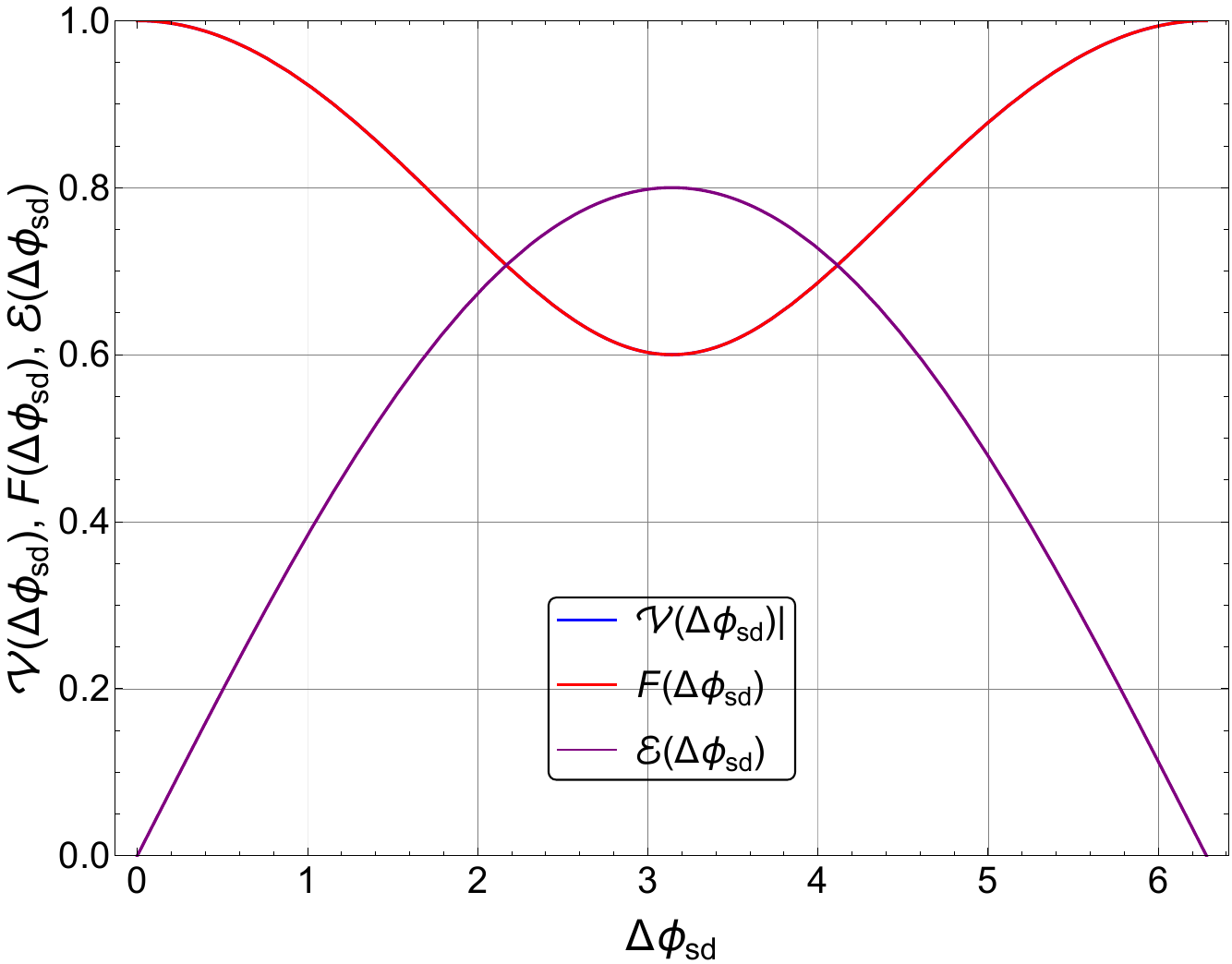}
\caption{\textbf{Phase-dependent coherence, fidelity, and entanglement.}
Coherence visibility \( \mathscr{V}(\Delta\phi_{sd}) \), fidelity \( F(\Delta\phi_{sd}) \), and entanglement measure \( \mathscr{E}(\Delta\phi_{sd}) \) are shown as functions of the seeding phase difference \( \Delta\phi_{sd} \). 
The fidelity determines the distinguishability of the conditional signal states and directly governs the visibility through \( \mathscr{V} = F \) for symmetric seeding. 
As the phase is varied, the fidelity decreases away from \( \Delta\phi_{sd} = 0 \), leading to reduced coherence visibility and increased entanglement. 
The results illustrate the complementary behavior between coherence and quantum correlations, satisfying the relation \( \mathscr{V}^2 + \mathscr{E}^2 = 1 \) for symmetric populations.}
\label{Fig5}
\end{figure}

\subsection{Coherence--entanglement trade-off}

Using the reduced-state formalism introduced above, the coherence visibility \( \mathscr{V} \) and entanglement measure \( \mathscr{E} \) can be expressed in terms of the reduced signal-mode density matrix. 
In particular, the entanglement is quantified through the purity of the reduced state,
\begin{equation}
\mu_s^2 = \mathrm{Tr}(\rho_{\mathrm{sig}}^2),
\end{equation}
such that
\begin{equation}
\mathscr{E}^2 = 1 - \mu_s^2,
\end{equation}
as established in Ref.~\cite{Yoon2021}. 
The coherence visibility \( \mathscr{V} \) is determined by the off-diagonal elements of the reduced density matrix and is directly related to the fidelity \(F\).

The coherence visibility \( \mathscr{V} \) is related to the fidelity \(F\) through
\begin{equation}
\mathscr{V} = \mathscr{V}_0 \, F,
\end{equation}
where \( \mathscr{V}_0 \) denotes the intrinsic visibility determined by the populations of the underlying states \cite{Yoon2021}. 
For symmetric seeding, \( \mathscr{V}_0 = 1 \), and the visibility reduces to \( \mathscr{V} = F \). 
Thus, the phase dependence of the fidelity directly determines the coherence properties of the signal mode.

Figure~\ref{Fig5} shows the phase dependence of \( \mathscr{V} \) and \( \mathscr{E} \) for symmetric seeding. 
As the phase difference \( \Delta\phi_{\mathrm{sd}} \) is varied, the fidelity changes, leading to a continuous redistribution between coherence and entanglement. The factorized structure $\mathcal{V}=\mathcal{V}_0 F$ is further confirmed
by the full parameter-space plots in Supplementary Figs.~S4 and S6.

For symmetric populations, this redistribution is governed by the complementarity relation
\begin{equation}
\mathscr{V}^2 + \mathscr{E}^2 = 1,
\end{equation}
which follows directly from the normalization of the reduced density matrix and its purity \cite{Yoon2021}. 
As \(F\) decreases, the off-diagonal coherence is reduced, leading to a decrease in \( \mathscr{V} \), while the mixedness of the reduced state increases, resulting in larger \( \mathscr{E} \). The complementary behavior between coherence and entanglement
is explicitly visualized in Supplementary Figs.~S5 and S6.

These results demonstrate that phase control provides a direct handle on the balance between coherence and quantum correlations in the ENBS system. 
Within the detector analogy, this behavior reflects how phase-dependent distinguishability of environmental states governs the trade-off between coherent interference and information encoded in the environment.

\subsection{Operational interpretation}

The above results establish that the ENBS platform enables controlled exploration of detector--environment correlations through phase-resolved measurements. 
The signal mode acts as an effective probe, while the idler mode provides a tunable environment whose properties determine observable coherence and correlation behavior.

While the system does not reproduce continuum quantum field dynamics, it captures the operational structure of detector response governed by environment-induced correlations in a fully controllable photonic setting.

\section{Conclusion and Outlook}

In summary, we have demonstrated that coherently seeded entangled nonlinear biphoton sources (ENBSs) provide a versatile and experimentally accessible platform for exploring phase-resolved detector--environment dynamics in a quantum-optical setting. 
By combining analytical modeling with realistic simulations, we have shown how signal photon number \( N_{\mathrm{sig}}(t) \), second-order correlations \( g^{(2)}(0;t) \), and phase-dependent interference measurements can be used to probe coherence and correlation properties of the system.

A central result of this work is that phase control enables continuous tuning of the distinguishability between conditional signal states, quantified by the fidelity \(F\), which in turn governs the balance between coherence and quantum correlations. 
The resulting behavior, captured through visibility and entanglement measures, provides a clear operational picture of how interference and distinguishability are linked in a controlled photonic environment.

Within this framework, the signal mode acts as an effective probe whose observable response is determined by correlations with an engineered environment formed by the idler mode and vacuum fluctuations. 
This establishes an operational analogy to detector--environment interactions in the Unruh--DeWitt (UDW) model, in which detector response is governed by correlation functions. 
At the same time, we emphasize that the present system does not reproduce continuum quantum field dynamics, causal structure, or thermal effects associated with accelerated detectors, and should therefore be understood as a controllable quantum-optical analog rather than a full relativistic field simulation.

Beyond its conceptual connection to detector models, the ENBS platform provides a flexible testbed for investigating phase-resolved quantum correlations and quantum information properties of photonic states. 
The ability to independently access intensity, first-order coherence, and second-order correlations offers a comprehensive experimental handle on the interplay between coherence, distinguishability, and quantum correlations.

Looking forward, this approach can be extended by incorporating multimode idler fields, spatially separated sources, or time-dependent control of pump and seeding parameters, enabling exploration of more complex correlation structures. 
Such extensions may provide routes toward simulating richer detector--environment scenarios and probing the role of structured environments in quantum dynamics.

Overall, our results establish coherently seeded ENBS systems as a powerful and scalable platform for studying phase-controlled quantum correlations and their connection to detector-based models in a fully accessible quantum-optical framework.

\bibliographystyle{apsrev4-2}
\bibliography{enbs_ref}

\section*{Methods}

The Supplementary Information (SI) provides full derivations of the theoretical model, including time-dependent expressions for the signal photon number \( N_{\rm sig}(t) \), the second-order correlation function \( g^{(2)}(0;t) \), and the reduced density matrix formalism used to evaluate fidelity, coherence visibility, and entanglement.

Here, we summarize the experimental architecture, modeling assumptions, and simulation parameters underlying the results presented in the main text.

\subsection*{Quantum optical platform}

The system consists of two coherently seeded single-photon frequency-comb (SPFC) sources implemented via type-0 spontaneous parametric down-conversion (SPDC) in periodically poled lithium niobate (PPLN) waveguides. 
Each source is pumped by a narrowband optical frequency comb centered at 530~nm and seeded by a continuous-wave idler field at 1542~nm.

The relative phases of the pump, seed, and signal paths are independently controlled, enabling implementation of the phase parameters \( \Delta\phi_p \), \( \Delta\phi_{\rm sd} \), and \( \Delta\phi_s \) introduced in the main text. 
These phases determine the effective interferometric phase governing the observed correlations.

\subsection*{Dynamical model}

The system dynamics are described by an effective Hamiltonian of the form
\begin{equation}
\hat{H}_{\text{ENBS}} = \hbar |g_{\text{eff}}| \sum_{j=1}^2 \left( e^{i\phi_{p,j}} \hat{A}_{\text{SM},j}^\dagger \hat{a}_{i,j}^\dagger + \mathrm{h.c.} \right),
\end{equation}
where \( \hat{A}_{\mathrm{SM},j} \) and \( \hat{a}_{i,j} \) denote signal and idler mode operators.

Optical loss is included via a Lindblad master equation,
\begin{align}
\frac{d\hat{\rho}}{dt}
=
-\frac{i}{\hbar}[\hat{H}_{\mathrm{ENBS}}, \hat{\rho}]
+
\sum_{j=1}^2 \left[ \mathcal{D}[\sqrt{\kappa_s}\hat{A}_{\mathrm{SM},j}] + \mathcal{D}[\sqrt{\kappa_i}\hat{a}_{i,j}] \right]\hat{\rho},
\end{align}
which is used for numerical simulations of dissipative dynamics. 
Analytical expressions in the main text are obtained under the same model in the weak-gain regime.

\subsection*{Regime of validity}

All analytical results are derived in the weak-gain regime, where the average photon number remains much smaller than unity and higher-order multiphoton processes are negligible. 
This ensures consistency with the reduced-state description used to define fidelity and coherence measures.

\subsection*{Numerical simulations}

Time-dependent observables, including \( N_{\mathrm{sig}}(t) \) and \( g^{(2)}(0;t) \), are obtained by solving the Lindblad master equation using truncated Fock-space representations. 
These simulations validate the analytical results within the parameter regime considered.

\subsection*{Experimental parameters}

Simulation parameters are chosen to match experimentally realistic conditions. 
The effective nonlinear coupling is set to \( |g_{\mathrm{eff}}|/(2\pi) \sim 1\text{--}10~\mathrm{GHz} \), with coherent seed amplitudes \( |\alpha| \) in the range 1--5. 
Signal-mode damping rates \( \kappa_s/(2\pi) \sim 2\text{--}3~\mathrm{GHz} \) correspond to optical loss over a 1-cm PPLN interaction length, with a group delay of approximately 75~ps.

\subsection*{Measurement observables}

The platform enables measurement of complementary observables: the signal photon number \( N_{\mathrm{sig}}(t) \), the second-order correlation function \( g^{(2)}(0;t) \), and the phase-dependent single-photon detection rate \( R(\Delta\phi_{\rm sd}) \), which provides direct access to first-order coherence. 
These measurements are consistent with established experimental implementations~\cite{Lee2018,Yoon2021}.

\bigskip
\noindent\textbf{Data availability} 
The data that support the findings of this study are available from the 
corresponding author upon reasonable request. 

\bigskip
\noindent\textbf{Code availability} The codes used for analysis of all figures are available from the corresponding author upon reasonable request.

\bigskip
\noindent\textbf{Acknowledgements} T. H. Yoon acknowledges Alexander Rothkopf for reading the manuscript and providing helpful comments, as well as Sun Kyung Lee and Mincheng Cho for their extended collaboration on SPFC state generation and its applications in the ENBS system. This work was supported by the National Research Foundation of Korea under Grant No. RS-2022-NR068815.

\bigskip
\noindent\textbf{Author contributions} T. H. Yoon conceived the research idea, developed the theoretical framework, performed all analytical derivations and numerical simulations, and wrote the manuscript. 

\bigskip
\noindent\textbf{Competing interests} The authors declare no competing interests.

\end{document}